\documentclass[journal]{IEEEtran}
\usepackage{cite}
\usepackage{siunitx}
\usepackage{comment}
\usepackage{amsmath,amssymb,amsfonts}
\usepackage{algorithmic}
\usepackage{graphicx}
\usepackage{textcomp}
\usepackage{url}

\def\BibTeX{{\rm B\kern-.05em{\sc i\kern-.025em b}\kern-.08em
    T\kern-.1667em\lower.7ex\hbox{E}\kern-.125emX}}

\begin{document}

\title{Ruggedized Embedded Ultrasound Sensing \\ in Harsh Conditions: eRTIS in the wild}

\author{
\IEEEauthorblockN{
    Dennis Laurijssen\IEEEauthorrefmark{1}\IEEEauthorrefmark{2},
    Wouter Jansen\IEEEauthorrefmark{1}\IEEEauthorrefmark{2},
    Arne Aerts\IEEEauthorrefmark{3},
    Walter Daems\IEEEauthorrefmark{1}\IEEEauthorrefmark{2},
    and Jan Steckel\IEEEauthorrefmark{1}\IEEEauthorrefmark{2}
     }\\
     \IEEEauthorblockA{\IEEEauthorrefmark{1}Cosys-Lab, University of Antwerp, Antwerp, Belgium}\\
    \IEEEauthorblockA{\IEEEauthorrefmark{2}Flanders Make Strategic Research Centre, Lommel, Belgium}\\
    \IEEEauthorblockA{\IEEEauthorrefmark{3}Service Group Materials, University of Antwerp, Antwerp, Belgium}\\

\thanks{Corresponding author: Jan Steckel (email: jan.steckel@uantwerpen.be)}
}

\maketitle

\begin{abstract}
We present eRTIS, a rugged, embedded ultrasound sensing system for use in harsh industrial environments. The system features a broadband capacitive transducer and a 32-element MEMS microphone array capable of 2D and 3D beamforming. A modular hardware architecture separates sensing and processing tasks: a high-performance microcontroller handles excitation signal generation and data acquisition, while an NVIDIA Jetson module performs GPU-accelerated signal processing. eRTIS supports external synchronization via a custom controller that powers and coordinates up to six devices, either simultaneously or in a defined sequence. Additional synchronization options include bidirectional triggering and in-band signal injection. A sealed, anodized aluminum enclosure with passive cooling and IP-rated connectors ensures reliability in challenging conditions. Performance is demonstrated in three field scenarios: harbor mooring, off-road robotics, and autonomous navigation in cluttered environments, demonstrates that eRTIS provides robust sensing in situations where optical systems degrade.
\end{abstract}

\begin{IEEEkeywords}
In-air ultrasound, Embedded systems, Microphone Arrays, Sound Source Localization, Acoustic signal processing, 3D Ultrasound, Ultrasound imaging, GPU acceleration
\end{IEEEkeywords}

\section{Introduction}
\label{sec:introduction}
In-air ultrasound sensing has gained attention as a robust alternative to optical systems for industrial applications, where conditions such as dust, fog, and rain can degrade camera- and LiDAR-based sensing~\cite{kerstens20193d,allevato2023ultrasonic,maier2021single,bogue2018applications}. Unlike commonly used optical methods, ultrasound exteroception remains resilient in such environments, making it highly suitable for use in harsh, real-world conditions. 

In-air ultrasound sensor arrays have been validated for several real-world applications. Object detection has been shown by A. Izquierdo et al. for the detection of humans from a moving vehicle in outdoor conditions~\cite{IZQUIERDO2024115586} or V. Madola et al. for detecting and localizing dairy cows~\cite{MADOLA2025110385}. X. Sun et al. created an ultrasound array and used a multi-view imaging technique for buried pipe inspections~\cite{SUN2025107762}. Some applications also use passive acoustic sensing for the detection of gas leaks~\cite{10950451, s24051366, 8956631} or UAVs~\cite{besnea2020acoustic,Blass2024,del2020spatial}, better known as drones.

Despite its advantages and its proven track record in several applications, practical deployment of ultrasound sensing in industrial settings in harsh conditions remains challenging. Indeed, real-time processing demands, mechanical robustness, and spatial resolution requirements place significant strain on conventional embedded systems. Many existing solutions would not be suitable for harsh real-world conditions outside of a lab or rely on external processing units~\cite{seckel2019physics,brauner20183}, which introduce latency and reduce system compactness.

To overcome these limitations, we present the eRTIS (Embedded Real-Time Imaging Sonar), a fully embedded, GPU-accelerated ultrasound sensing system tailored for harsh industrial environments. This sensor system, of which the underlying principles were first presented in~\cite{kerstens2019}, incorporates a broadband ultrasound emitter and exhibits an improved spatial resolution compared to narrow-band alternatives by enabling broadband pulse excitation and frequency-domain analysis. Furthermore, a random sparse microphone array is used to perform 3D-beamforming, allowing spatially aware sensing and localization in three dimensions without mechanical scanning~\cite{steckel2012broadband}.

At the core of the system lies the combination of a microcontroller-based data acquisition/signal conditioning system and an embedded NVIDIA Jetson module that combines a multi-core CPU with a CUDA-capable GPU. This heterogeneous computing platform allows us to implement computationally demanding tasks—such as beamforming, signal conditioning, and envelope detection—directly at the edge. By performing all processing locally, the system minimizes latency and reduces the need for high-bandwidth data transmission. All combined, eRTIS enables real-time performance for data acquisition, signal-processing, and downstream applications.  

Previous works in airborne ultrasound sensing have typically focused on narrowband piezo-electrical transducers~\cite{9126862, 9593561, rutsch2021waveguide, konetzke2015phased, allevato2023two, allevato2022air, allevato2020real}. Typically, these narrowband transducers are coupled with a small-channel-count microphone array, decreasing system resilience to noise and spatial resolution. In the literature, several 2D-array geometries have been presented but often depend on external processing setups or lack resilience for field deployment. For a more complete list of references and review on in-air sonar sensing, we refer the reader to~\cite{kerstens2020advanced}. To the best of our knowledge, this is one of the first implementations of a fully embedded, broadband, 3D in-air ultrasound sensor built specifically for industrial and harsh environments.

This paper presents the full design and implementation of eRTIS, detailing its sensing hardware, embedded data acquisition architecture, and GPU-accelerated signal processing pipeline. We envision this paper as a guideline of aspects that need to be considered when developing an industrially viable 3D-sonar sensor which can operate in harsh environments, while being flexible in its application. 

In addition to sharing the design philosophy, we evaluate the system’s performance under representative real-world conditions, demonstrating that ultrasound—when combined with embedded GPU acceleration—can serve as a viable and robust alternative to traditional optical sensing methods.

To validate the sensing capabilities of eRTIS, we conducted experiments across three distinct real-world application scenarios:

    \begin{enumerate}
        \item Harbor mooring vessel monitoring, where eRTIS was used to detect and localize large structures in marine environments, demonstrating robustness against fog, mist, and reflective surfaces.
        \item Outdoor off-road robotics, showcasing the system’s ability to operate reliably in dusty and unstructured terrain where cameras and LiDAR are prone to failure.
        \item Autonomous Guided Vehicle (AGV) navigation in cluttered indoor environments, where eRTIS enabled real-time spatial awareness and obstacle detection for guided vehicles operating among densely packed machinery and infrastructure.
    \end{enumerate}

These case studies highlight the system’s versatility, resilience, and potential to complement optical sensing solutions in industrial and autonomous applications.

\section{Hardware Architecture}
\label{sec:hardware_architecture}
The eRTIS system is designed with a strong emphasis on modularity, ruggedness, and industrial integration. The hardware is divided into a front-end (transceiver subsystem) and back-end (embedded data acquisition and processing subsystems) PCB, allowing for clear separation between sensing components and processing/control logic. This modular approach not only simplifies development and testing but also enables future upgrades or adaptations for different sensing configurations and industrial requirements.

To ensure robust operation in harsh environments, the system is housed in a custom-designed enclosure that fulfills multiple roles. Mechanically, it serves to ruggedize the system, protecting it from mechanical shock, vibration, and environmental ingress. Thermally, the enclosure doubles as a passive heat sink, drawing heat away from the NVIDIA Jetson module and maintaining stable operation under high processing loads. The enclosure has been designed with water and dust proofing in mind, integrating appropriate sealing features to comply with industrial protection standards (e.g., IP-rated interfaces).

All I/O interfaces—including power, communication, and synchronization lines—are equipped with environmentally sealed connectors to maintain water and dust resistance without compromising connectivity. This enables reliable integration with external systems in a wide range of industrial settings, from factory floors to outdoor deployments.

To support distributed sensor networks or coordinated multi-sensor operation, eRTIS includes provisions for external synchronization. An external sync input allows the system to be triggered or aligned with other sensors or controllers, enabling precise timing control across systems.

Together, these hardware features make eRTIS well-suited for embedded deployment in environments where traditional, lab-grade sensing systems would struggle to operate reliably.

\subsection{Transceiver Subsystem}
\label{subsec:tranceiver_subsystem}
The front-end PCB of the eRTIS system houses the complete ultrasound transceiver subsystem, responsible for both emitting and receiving acoustic signals in air. This board is designed with modularity and compactness in mind, balancing high-voltage operation with sensitive signal reception and precise timing coordination.

The transmitter section features a high-voltage bias generator and a high-voltage amplifier with two gain stages, which together drive a SensComp 7000 capacitive ultrasound transducer. This air-coupled capacitive transducer operates over a broad frequency range, making it well-suited for broadband excitation. eRTIS uses logarithmic frequency-modulated (FM) sweeps between 20~\si{\kilo\hertz} and 80~\si{\kilo\hertz} as the excitation signal. These log-FM chirps offer high time-bandwidth products and enhanced Doppler tolerance, enabling reliable operation in scenarios involving relative motion—such as moving targets or dynamic industrial environments.

For reception, the front-end integrates a microphone array composed of up to 32 Syntiant, formerly known as Knowles, SPH0641LU4H-1 digital PDM microphones. This array allows for high-resolution spatial sampling and supports two distinct configurations:
\begin{itemize}
    \item A 6×5 regular grid, which allows spatial smoothing~\cite{verellen2020,verellen2020urtis} due to the regular sensor geometry.
    \item A randomized array~\cite{steckel2012broadband} based on Poisson disc sampling, which minimizes grating lobes and improves angular resolution in wideband beamforming applications.
\end{itemize}

This flexible array topology allows the system to be tailored to the spatial resolution and field-of-view requirements of specific use cases.

Each PDM microphone is connected to the back-end data acquisition subsystem via synchronized I/O lines. Special attention is given to clock distribution and timing alignment, ensuring phase coherence across all microphone channels—an essential prerequisite for accurate beamforming and time-of-flight-based imaging. Our 2D planar array configuration enables the resolution of sound sources in 3D space by beamforming into directions of interest in both the azimuth and elevation vectors.

\begin{figure*}[t!]
    \centering
    \includegraphics[width=1\linewidth]{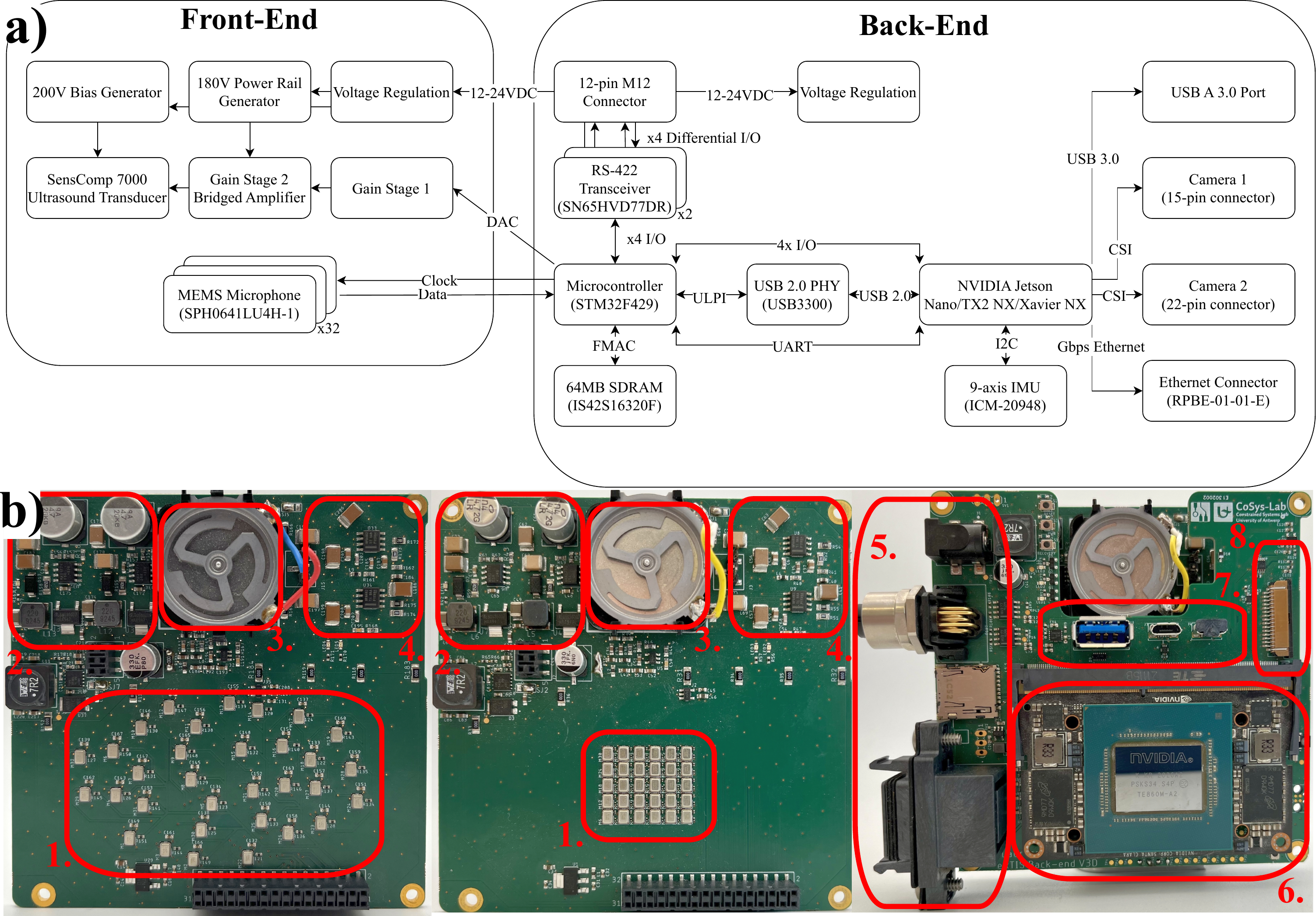}
    \caption{a) Shows the schematic representation of the hardware architecture showing the front and back-end with their respective subsystems and interconnections. b) The eRTIS sensor printed circuit boards with the randomized 32-element microphone array on the left and the 6×5 regular grid front-end in the middle. The back-end PCB is shown on the right with an NVIDIA Xavier NX system on module inserted in the DDR4 socket. In this sub-figure the following components are highlighted 1. the 32-element MEMS microphone array 2. the 180~\si{\volt} and 200~\si{\volt} step-up regulation circuits 3. the Senscomp 7000 ultrasound transducer 4. the two-stage gain amplifiers that amplify the output signal 5. the connectors for providing power and interfacing with the sensor 6. the NVIDIA System on Module 7. USB connectors for providing extra data storage capabilities or directly interfacing with the microcontroller 8. one of two CSI connectors that provide a means for connecting a camera directly with the NVIDIA SoM. }
    \label{fig:hw_data_flow}
\end{figure*}

\subsection{Embedded Data Acquisition Subsystem}
\label{subsec:embedded_data_aquisition_subsystem}
The back-end PCB of the eRTIS system houses the embedded data acquisition subsystem, which is responsible for signal generation, microphone clocking, synchronized data capture, and communication with the main processing unit. At the core of this subsystem is an STM32F429 microcontroller, selected for its high-speed peripherals, memory interface, and advanced timing capabilities.

One of the unique features of this implementation is the use of the STM32F429’s integrated digital-to-analog converter (DAC) to generate the excitation signal for the capacitive ultrasound transducer. This approach enables flexible, runtime-adjustable waveform generation, allowing for experimentation with different excitation profiles, including a logarithmic FM sweep or constant sine waves.

For the receiver path, the STM32F429 generates the PDM clock signal for the microphone array and synchronously captures the digital data streams from up to 32 SPH0641LU4H-1 microphones. The captured data is streamed through Direct Memory Access (DMA) into a 64 MB external SDRAM (IS42S16320F-7BL), which is accessed via the microcontroller’s Flexible Memory Controller (FMC) interface. This memory provides enough buffer space to store high-resolution snapshots of the incoming acoustic wave fronts.

High-speed communication between the STM32F429 and the embedded NVIDIA Jetson module is established via a USB 2.0 high-speed interface, implemented using the STM32’s ULPI peripheral in conjunction with an external USB3300 PHY from Microchip. This configuration enables reliable bulk data transfer with low latency, ensuring the Jetson module has continuous access to incoming data for further GPU-accelerated processing.

In addition to USB, the system includes differential communication I/O via two SN65HVD77DR RS-422 transceivers through a 12-pin dust and waterproof-shielded M12 connector. This interface is designed for long-distance, noise-immune communication, making it suitable for connecting eRTIS to external controllers or triggering devices in electrically noisy industrial environments.

By offloading time-critical acquisition and control tasks to the STM32F429, the system ensures low-latency and deterministic operation while enabling the Jetson module to focus on compute-intensive processing and data interpretation.

\subsection{Embedded Processing Subsystem}
\label{subsec:embedded_processing_subsystem}
The embedded processing subsystem in eRTIS is built around a modular carrier platform designed to support a range of NVIDIA Jetson system-on-modules, including the Jetson Nano, Jetson TX2 NX, Jetson Xavier NX, Jetson Orin Nano and Jetson AGX Orin. This modularity enables scalability depending on the computational needs—ranging from lightweight inference tasks to intensive real-time 3D signal processing leveraging CUDA-enabled GPU acceleration.

The STM32F429-based acquisition subsystem communicates with the Jetson module via a USB 2.0 high-speed interface, implemented using the ULPI peripheral and an external USB PHY. While this link supports transfer of buffered measurement data, it also serves as a bidirectional control interface. Through this channel, the Jetson module can dynamically configure measurement parameters, adjust acquisition settings, and modify the excitation waveform used for the transducer in real time—enabling adaptive operation and experiment reconfiguration without physical access to the system.

To facilitate integration into larger networks or remote monitoring setups, the Jetson module is equipped with a Gigabit Ethernet interface, using a Samtec RPBE-01-01-E connector to ensure environmental sealing against water and dust ingress. This connection supports fast data offloading, remote access through SSH or even RDP, and synchronization with industrial control systems or distributed sensor networks.

To enhance environmental awareness and enable correction for sensor orientation, a TDK Invensense ICM-20948 9-axis IMU is connected via I2C. This sensor provides accelerometer, gyroscope, and magnetometer data, allowing eRTIS to track its motion and orientation during measurements—particularly useful when mounted on mobile robot platforms or robotic arms.

The system also includes several expansion interfaces:

    \begin{itemize}
        \item Dual two-lane CSI camera inputs are compatible with both 15-pin and 22-pin camera modules from various manufacturers, allowing synchronized optical/acoustic data capture.
        \item An SD card slot for extended local storage, useful for logging large volumes of raw or processed data.
        \item A USB 2.0 host port for connecting peripheral devices such as USB cameras, Wi-Fi dongles, Ethernet adapters, or external USB drives, enabling further system expansion and adaptability to diverse deployment scenarios.
    \end{itemize}

Together, these capabilities make the embedded processing subsystem a powerful and adaptable platform for real-time acoustic imaging in challenging industrial environments.

\subsection{Enclosure for Harsh Environments}
\label{subsec:enclosure}
\begin{figure}
    \centering
    \includegraphics[width=1\linewidth]{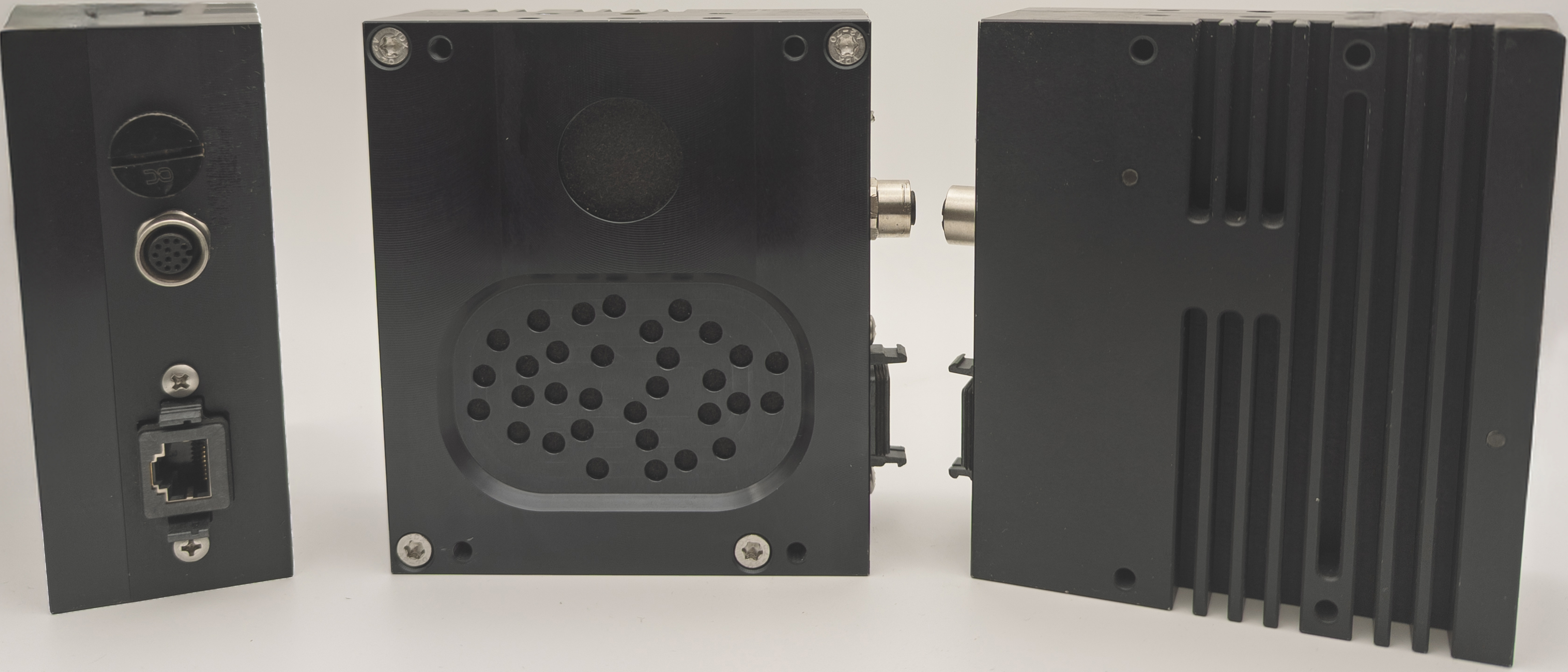}
    \caption{The eRTIS anodized aluminum case is shown displaying it's ingress proof connectors on the left side that provide power and interfaces for data transfers and triggering the sensor. In the middle the front of the case is showing the black mesh membrane through the cutouts in the enclosure and on the right is the back of the case showing the cooling slots and two status light indicators.}
    \label{fig:caseSides}
\end{figure}

To ensure reliable operation in harsh industrial environments, the entire eRTIS system is enclosed in a custom-designed, anodized aluminum housing shown in Fig.~\ref{fig:caseSides} that serves both mechanical and thermal functions. The enclosure is CNC-machined from solid aluminum for high structural rigidity, protecting sensitive electronics from vibration, shock, and environmental exposure. Its anodized surface finish enhances corrosion resistance and electrical insulation, making it suitable for long-term deployment in marine, outdoor, and industrial environments.

Thermal management is addressed through direct thermal coupling between the enclosure and heat-generating components, particularly the NVIDIA Jetson module. The enclosure acts as a passive heat sink, dissipating heat through conduction and natural convection, thereby ensuring stable operation under continuous compute loads without the need for active cooling mechanisms like fans—which are often unreliable in dusty or humid environments.

\begin{figure*}
    \centering
    \includegraphics[width=1\linewidth]{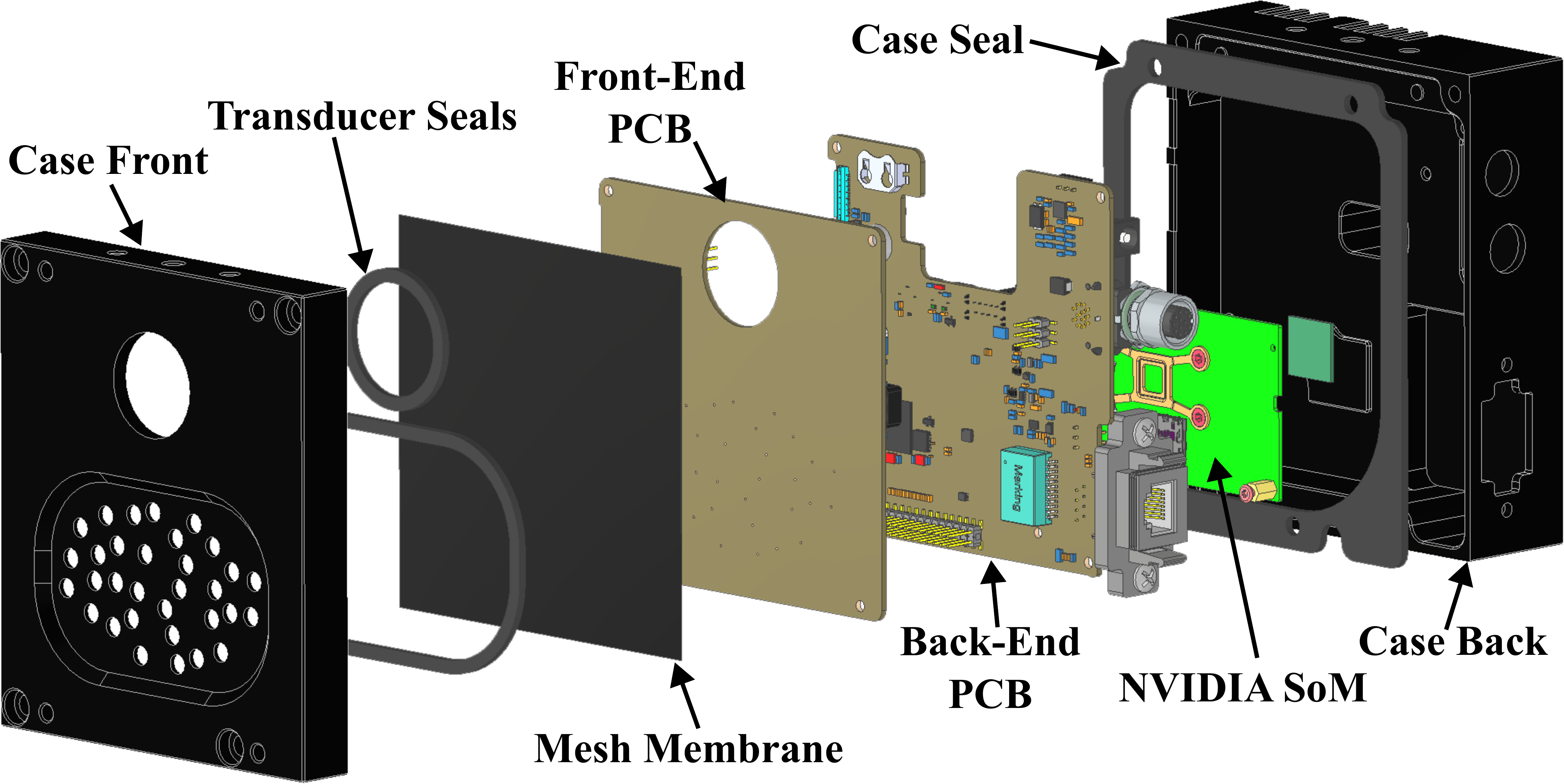}
    \caption{An exploded view of the CAD model of the eRTIS sensor and its anodized aluminum enclosure together with its seals and mesh membrane for dust and water ingress protection.}
    \label{fig:explodedview}
\end{figure*}

\begin{figure*}
    \centering
    \includegraphics[width=1\linewidth]{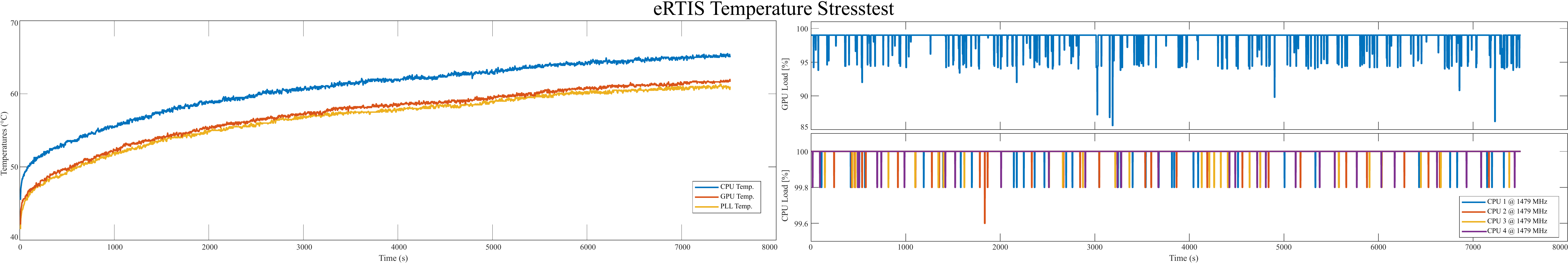}
    \caption{In order to verify the heat dissipating properties of the designed enclosure, a stress test was devised where the device was placed in an ambient office environment and both the CPU and GPU of the NVIDIA Jetson Nano were put under a constant synthetic load for over 2 hours of constant operation. The left plot shows the temperature of the CPU, GPU and PLL of the NVIDIA Jetson Nano whereas the two plots on the right show the respective load of the GPU and CPU cores.}
    \label{fig:stress}
\end{figure*}

To evaluate the thermal performance of the enclosure, the NVIDIA Jetson Nano was subjected to a sustained stress test under ambient office conditions. The CPU load was generated using the Python package stress, while the GPU load was produced by a custom Python beamforming workload implemented with CUDA. This ensured that both processing units were driven with realistic, high-intensity workloads representative of the intended application domain. The system was operated continuously for over two hours, during which the temperatures of the CPU, GPU, and PLL were monitored alongside the utilization of the GPU and individual CPU cores. The results confirm that the enclosure performs as intended, as the maximum measured temperature during the test did not exceed \ang{66}~C, as shown in Fig.~\ref{fig:stress}.

To meet IP65 or higher ingress protection, the enclosure uses custom-cut environmental seals at the mating interface between its two halves. These seals are fabricated to precisely match the enclosure geometry and provide a tight, water- and dust-resistant barrier. All cable entries and ports are sealed using ruggedized connectors rated for harsh environments, further preserving the integrity of the housing. An exploded view of the sensor with its enclosure is shown in Fig.~\ref{fig:explodedview} where all of the internal parts can be seen.

A shielded 12-pin M12 connector provides both power delivery and access to differential I/O lines directly interfaced with the STM32F429 microcontroller. These include RS-422-compatible communication lines and a synchronization input, enabling noise-resistant signaling and integration with external control systems or sensor networks. The M12 connector ensures secure locking and robust Electromagnetic interference (EMI) shielding.

For high-speed data connectivity and remote system control, a Gigabit Ethernet interface is included via a Samtec RPBE-01-01-E sealed Ethernet connector, offering reliable, IP-rated network access.

\begin{figure}
    \centering
    \includegraphics[width=1\linewidth]{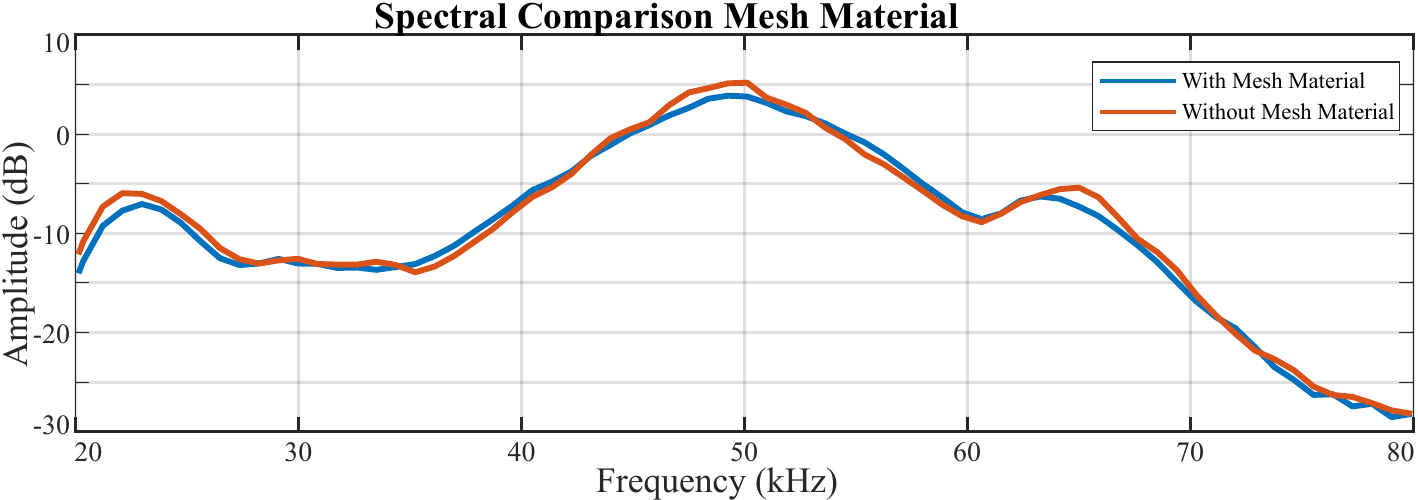}
    \caption{Comparison of the received ultrasound frequency spectrum with and without the protective mesh membrane.}
    \label{fig:gore-comparison}
\end{figure}

To protect the sensing components without compromising performance, the microphone array and ultrasound transducer are shielded behind an acoustic protective mesh material produced by Gore. This mesh is acoustically transparent for both audible and ultrasound signals while acting as a barrier against water, dust, and debris, preserving sensor sensitivity and array functionality in the field. To assess the attenuation of the mesh material on the acoustic signals, experiments were performed with the eRTIS device emitting broadband calls and measuring their reflection. The results showed that there is practically no attenuation on the received signals when comparing it to the results without a mesh membrane, as can be seen in Fig.~\ref{fig:gore-comparison}.

\newpage
\noindent Together, the enclosure’s mechanical, thermal, and environmental protection features—validated in laboratory tests simulating outdoor conditions such as fog, rain, and dust—ensure that eRTIS remains reliable and fully operational in challenging real-world environments, including maritime, off-road, and industrial scenarios.

\subsection{Synchronization}
\label{subsec:synchronization}
To support precise temporal coordination and enable multi-sensor deployments, eRTIS incorporates a versatile synchronization and timing interface that supports both external and internal synchronization modes. These capabilities are essential for applications such as multi-node sensing arrays, data fusion with other modalities, and synchronized measurements in dynamic environments.

The STM32F429 microcontroller manages all time-critical tasks, including waveform generation, microphone sampling, and external trigger handling. A dedicated differential synchronization input, exposed through the shielded 12-pin M12 connector, provides a noise-immune RS-422-compatible line for receiving external timing signals. This low-latency interface enables the system to react with minimal delay to incoming synchronization triggers.

A custom-built external synchronization unit (hereafter referred to as a multi-system trigger device) is available to interface with up to six eRTIS devices simultaneously. This unit not only supplies power but also distributes a synchronization signal to each device. It supports two operational modes:
\begin{itemize}
  \item Simultaneous triggering, where all connected devices start their acquisition cycles at the same time.
  \item Sequenced triggering, where each device receives its trigger in a predetermined order, enabling non-overlapping or staggered operation in cases where a shared sync signal is not feasible or desirable.
\end{itemize}

In addition to receiving sync signals, eRTIS can also generate them, making it suitable for bidirectional synchronization. It can output a trigger pulse or activate a visible LED in the Field of View (FOV) of camera systems to align acoustic and visual data~\cite{10.1242/jeb.173724} streams—useful for SLAM, mapping, or inspection workflows involving external vision systems.

Furthermore, the system supports in-band synchronization, wherein a synchronization signal—either a distinct pulse or a continuous waveform—can be injected directly into the microphone data stream. This allows precise alignment of recorded data during post-processing, even in setups without dedicated sync lines. The injected signature can be uniquely identified in the recorded acoustic data, providing a reliable reference point for time-stamping and frame alignment.

This combination of hardware-level trigger support, external sync unit integration, and software-assisted in-band markers makes eRTIS highly adaptable to a wide range of sensing and coordination strategies across diverse application domains.

\section{Software Architecture}
\label{sec:software_architecture}
The eRTIS sensor has a microcontroller and an embedded processing subsystem. Both run specific low- and high-level software, respectively, to get to the result of real-time acoustic images and downstream applications. In this section we will show how these images are created using GPU-accelerated digital signal processing and the software packages created for the various application types where these sensors can be deployed. 

\begin{figure*}[t!]
    \centering
    \includegraphics[width=1\linewidth]{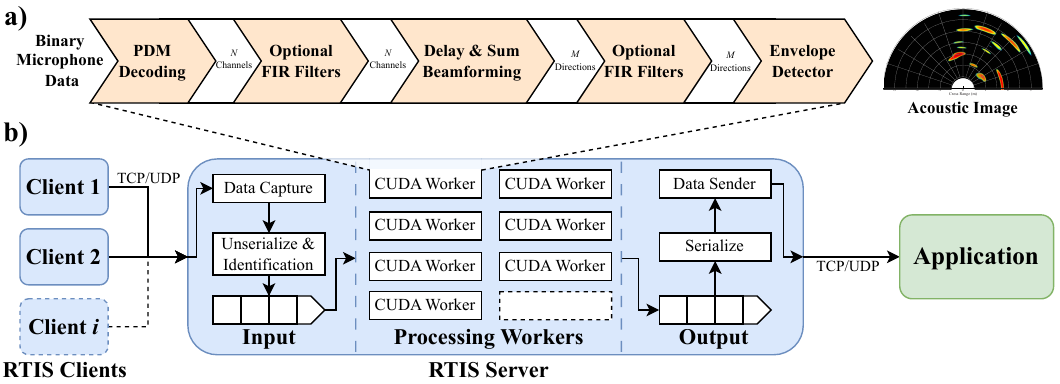}
    \caption{a) Flow diagram showing the steps going from the PDM encoded microphone signals to acoustic images using the GPU-accelerated \textit{RTIS CUDA} signal processing. Optional filters are, for example, a matched filter when using active sensing. Afterward, a beamformer generates a spatial filter for every direction of interest. The output is then put through an envelope detector for every direction. b) Diagram showing the \textit{RTIS Network} data flow from the clients to the application. The sensor nodes (clients) send their data packages to the central node (server). This node will identify each package and add it to a queue. The server node has a series of multi-threaded workers which will pick from that queue and process the measurement using \textit{RTIS CUDA}. These resulting acoustic images will be put in another queue to be sent to the application.}
    \label{fig:sw_data_flow}
\end{figure*}

\subsection{Data Processing Flow}
\label{subsec:data_processing_flow}
The signals of the microphone array on the eRTIS are stored on 64 MB external SDRAM in the embedded data acquisition subsystem by the STM32 microcontroller. They are encoded using PDM modulation to binary data. First, this data is transferred over a serial connection to the embedded processing subsystem. Afterwards, a Python C++ module called \textit{RTIS CUDA} will use GPU acceleration on the NVIDIA GPU to convert this binary data into the acoustic images in real time. All 1D and 2D filters used in this DSP pipeline are implemented as FFT convolution filters. For this, we can make use of the \textit{CuFFT} CUDA library for paralleled (i)FFT execution. A more in-depth description of the GPU-accelerated signal-processing can be found in~\cite{jansengpuprocessingnetwork}. 

The encoded data is first decoded using a low-pass filter back into the microphone signals. Next, when performing active sensing using the emitter, the broadband FM sweep and its reflected echoes are extracted from the noise using an optional matched filter.

After this, a conventional beamforming algorithm in the time-domain takes place. For each direction of interest, we calculate the delay relative to each microphone channel. This allows for steering in directions of interest with simple delay-and-sum beamforming~\cite{VanTrees2002OptimumProcessing}. This step is sped up using a custom GPU kernel with atomic operations.

Finally, after conventional beamforming, we can extract the envelope of the signal using an additional filter. The result is the reflector distribution for each particular direction. We now have a 2D or 3D image that gives us the intensity, distance, and direction as dimensions. These digital signal processing steps used are shown in Fig.~\ref{fig:sw_data_flow}a. 

In most applications using our acoustic images we usually also introduce additional adaptive filtering and noise reduction steps to further improve the results in harsh environments to compensate for non-linear unpredictable noise introduced by the environment. 

\begin{figure}
    \centering
    \includegraphics[width=1\linewidth]{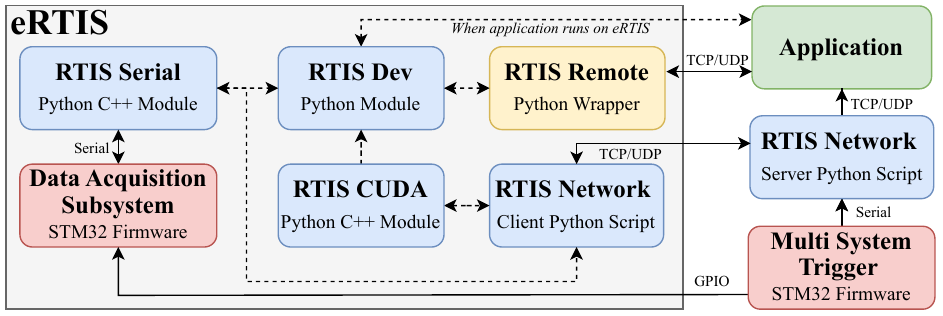}
    \caption{Diagram showing the complete software architecture that runs on the embedded processing subsystem as well as on external systems. The dotted lines represent the usage of Python APIs for the implementation. It shows the interaction between the low-level firmware on the STM32 microcontroller, the various Python modules, and end-user applications. The two main variations of using a single sensor using \textit{RTIS Dev} or multiple synchronized sensors using \textit{RTIS Network} with the multi-system trigger device are shown.}
\label{fig:software_architecture}
\end{figure}

\subsection{API Interface}
\label{subsec:api_interface}
To allow for a user-friendly experience for operating the eRTIS sensor, the high-level software is divided into several modular packages that can interact depending on the application. The microphone data is captured and stored in memory by the STM32 microcontroller on the embedded data acquisition subsystem. A Python C++ Extension module called \textit{RTIS Serial} was made to have a fast communication protocol between the microcontroller and the embedded processing subsystem over USB. \textit{RTIS Serial} can send commands to the STM32 microcontroller to configure its capture settings, as well as pull the latest microphone signals from the on-board memory. As explained in~\ref{subsec:data_processing_flow}, \textit{RTIS CUDA} can use the encoded data and process it in real-time to acoustic images.

For general usage applications requiring acoustic sensing, a single sensor is used. In these cases, a Python module called \textit{RTIS Dev} is used and available on the embedded processing subsystem that uses \textit{RTIS Serial} and \textit{RTIS CUDA} to allow the user to quickly configure capture and processing with the eRTIS sensor. However, \textit{RTIS Dev} is installed on the eRTIS itself, meaning that when no direct interactive interface is available to the user, it is cumbersome to operate. Therefore, an IP-based wrapper called \textit{RTIS Remote} is available that can be used from an external device in the same LAN as the eRTIS device to access the \textit{RTIS Dev} API functions remotely over a TCP connection. 

If multiple eRTIS sensors are used as described in~\ref{subsec:synchronization} with the multi-system trigger device, instead of \textit{RTIS DEV}, the \textit{RTIS Network} framework will be used. A more in-depth description can be found in~\cite{jansengpuprocessingnetwork}. The communication layer between the various sensor nodes and the central server node is a straightforward client-server communication layer over IP (TCP/UDP). The client side will serialize the data combined with an identification serial number and a timestamp. Afterward, they send this data package over TCP to the server. There, each package is unserialized, identified, and added to a queue. Either the client will take care of the (partial) signal processing using \textit{RTIS CUDA} on-device or the server can perform this step (partially). The server module can have a series of multi-threaded processes defined as CUDA workers which will pick from that queue and process them. This communication layer is shown in Fig.~\ref{fig:sw_data_flow}b. The server node will also take care of the configuration of the clients for capture and processing. It has a Web GUI to interact with as well as a REST and CLI API. A diagram with an overview of all mentioned software packages and how they interact can be found in Fig.~\ref{fig:software_architecture}.

Several real-time applications have been created and demonstrated in previous publications that have used this software architecture. From single sensor applications for detection, classification, and tracking~\cite{blankers2024fusion,de2023detecting,kerstens2023tracking,jansen2024cnnlandmarks} to a multi-sensor application for navigation~\cite{jansen2022real}. It has also facilitated research into electronic acoustic image stabilization~\cite{jansen2024steadirtis}.


\section{Experimental Setup and Results}
\label{sec:results}

\subsection{Real-Time Processing}
\label{subsec:real-time-processing}

\begin{figure}
    \centering
    \includegraphics[width=1\linewidth]{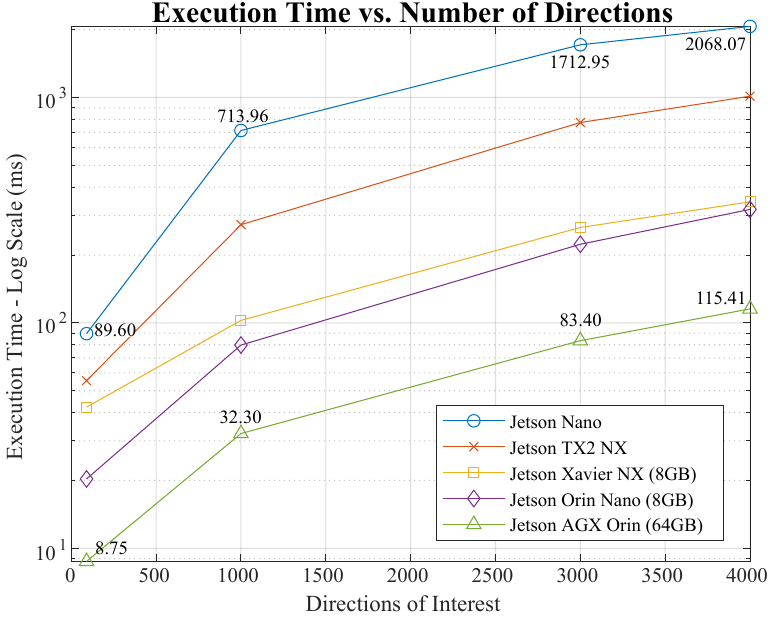}
    \caption{Execution times of the GPU-accelerated \textit{RTIS CUDA} signal processing from binary microphone data to acoustic images for some common resolutions beamforming indicated by the number of directions of interest (90, 1000, 3000 \& 4000). A matched filter and envelope detector were also enabled. Different NVIDIA Jetson platforms are compared as the embedded processing subsystem. It shows the mean execution time over 50 runs for the five platforms. The largest standard deviation measured was on the Jetson Nano platform for 3000 directions at 34.07 ms. All used a capture length of 163,840 samples, which corresponds to a recording time of 36.4 ms.}
    \label{fig:gpu_timings}
\end{figure}

Over the last few years, NVIDIA has released multiple versions of the Jetson platform, ranging from the original Jetson Nano to the recent Jetson AGX Orin, with large performance increases each year. To scale the available compute power and memory availability depending on the application and task, we have used five Jetson platforms so far. 

\noindent With an increase in CUDA cores, one can observe a faster processing time that increases linearly with the processing resolution of the acoustic images when beamforming. A comparison of the various NVIDIA Jetson platforms and processing resolutions can be found in Fig.~\ref{fig:gpu_timings}. On the latest Jetson version, with 4000 directions, it is nearing a 10 Hz processing rate. 

\subsection{Accuracy \& Resolution Validation}
\label{subsec:accuracy_resolution_validation}
\begin{figure*}
    \centering
    \includegraphics[width=1\linewidth]{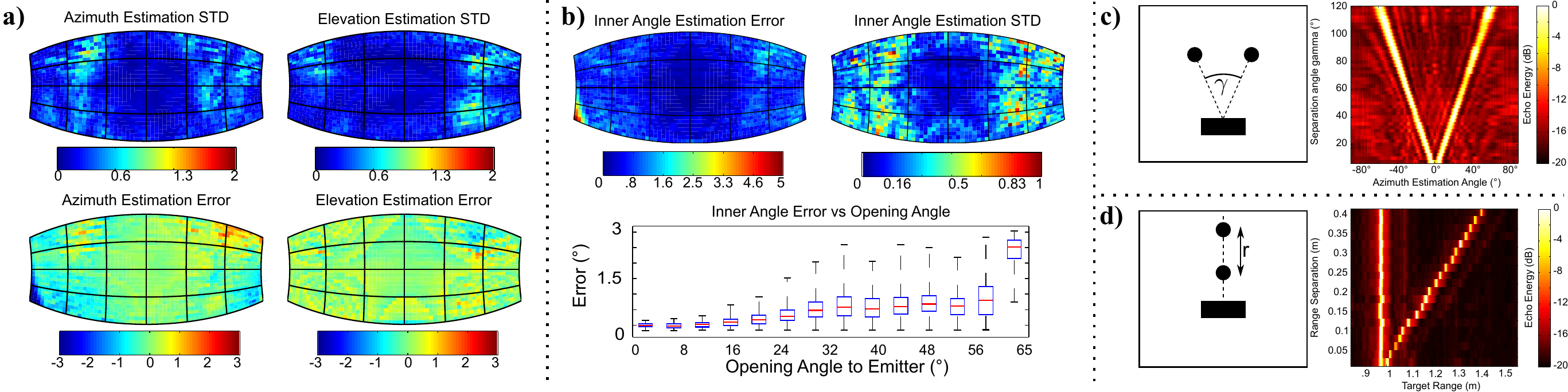}
    \caption{Demonstration of the spatial accuracy and resolution of the eRTIS sonar sensor in 3D.  Panel a) shows the estimated azimuth and elevation errors when localizing a a single spherical target (diameter: 8 cm), as well as the standard deviation on the location estimates. The plots are a Lambert Equal Area projection, with grid lines spaced 20° in azimuth and 14° in elevation. Panel b) shows the localization errors expressed in inner angle with respect to the sensor’s normal direction, providing a compact representation of the 3D localization accuracy. Panel c) shows experimental results for two identical poles placed at identical ranges, producing simultaneous echoes for varying inner angles, illustrating the system’s angular resolution and capability of resolving simultaneously arriving echoes. Panel d) shows results for two poles positioned at identical directions in front of the sensor but at varying range separations, demonstrating the range resolution capability of the eRTIS system.}
\label{fig:quantResults}
\end{figure*}

To provide a complete context for the real-world results presented in the next sub-section, it is important to first establish the validated performance envelope of the sensor solution. The fundamental accuracy, resolution, and robustness of the array system have been rigorously analyzed and documented in our prior publications \cite{6385584, 6331017}. This body of work demonstrated the array's capability to achieve high spatial resolution alongside metric accuracy under varying operational conditions. To summarize and visualize these foundational performance metrics, we present a combined synthesis of key results from our previous studies in Fig.~\ref{fig:quantResults}.

\subsection{Real-World Experiments (in the Wild)}
\label{subsec:real_world_experiments}
\begin{figure*}
    \centering
    \includegraphics[width=1\linewidth]{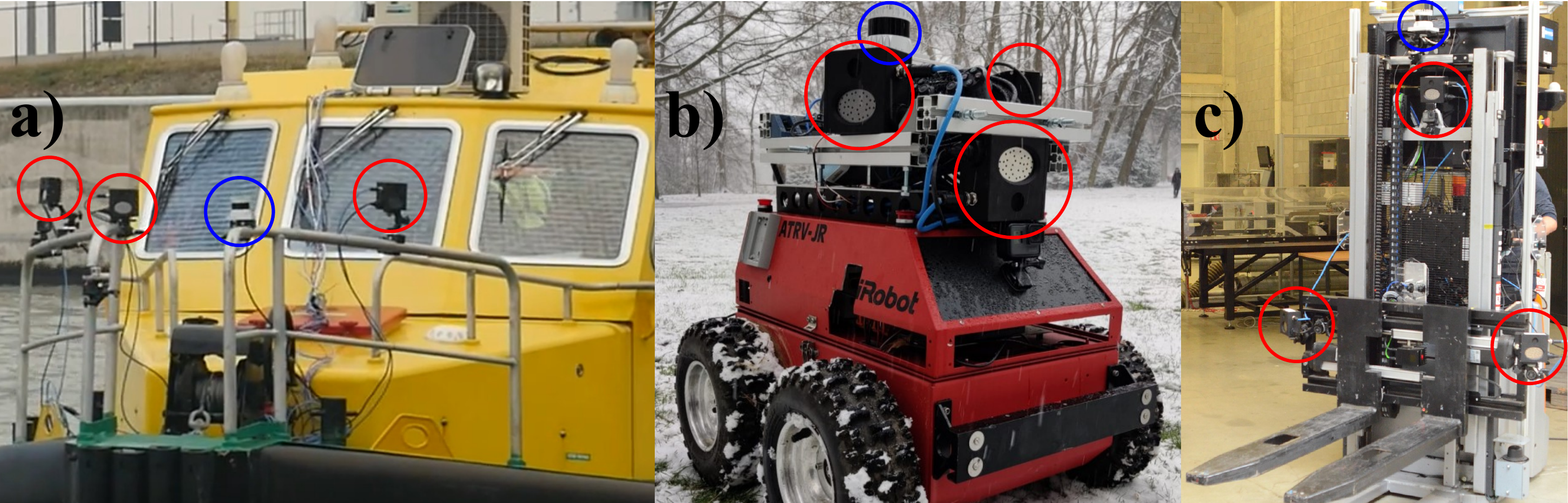}
    \caption{Pictures of the real-world experiment setups with three eRTIS sensors (highlighed in red) and the Ouster LiDAR (highlighted in blue) were the sensors are: (a) Mounted on a harbor mooring vessel in the Port of Antwerp. (b) Mounted on an outdoor robot. (c) Mounted on forklift in an industrial setting with glass safety barriers.}
\label{fig:expSetups}
\end{figure*}

\color{black}
To validate the system’s ability to operate \emph{in the wild}, next to the lab ingress tests, several domain-specific experiments were conducted in outdoor environments representative of maritime, off-road, and industrial use cases. These tests employed the software architecture described in Section~\ref{sec:software_architecture}, utilizing a multi-sensor configuration with three eRTIS units synchronized through the dedicated synchronization module. Furthermore, the datasets also included a single Ouster OS0-128 LiDAR pointcloud and a camera image for comparison. The used application connected to the \textit{RTIS Network} framework was a Robot Operating System (ROS) node~\cite{ros} that will publish all sensor data captured by these sensors. Fig.~\ref{fig:expSetups} shows the sensor setups that were used in these three applications, were the eRTIS sensors are highlighted with red circles and the Ouster LiDAR with a blue circle. In these experiments there is some overlap of field-of-view between the eRTIS sensors to be more resilient in harsh-environments and to have some form of physical redundancy. 

\begin{figure*}
    \centering
    \includegraphics[width=1\linewidth]{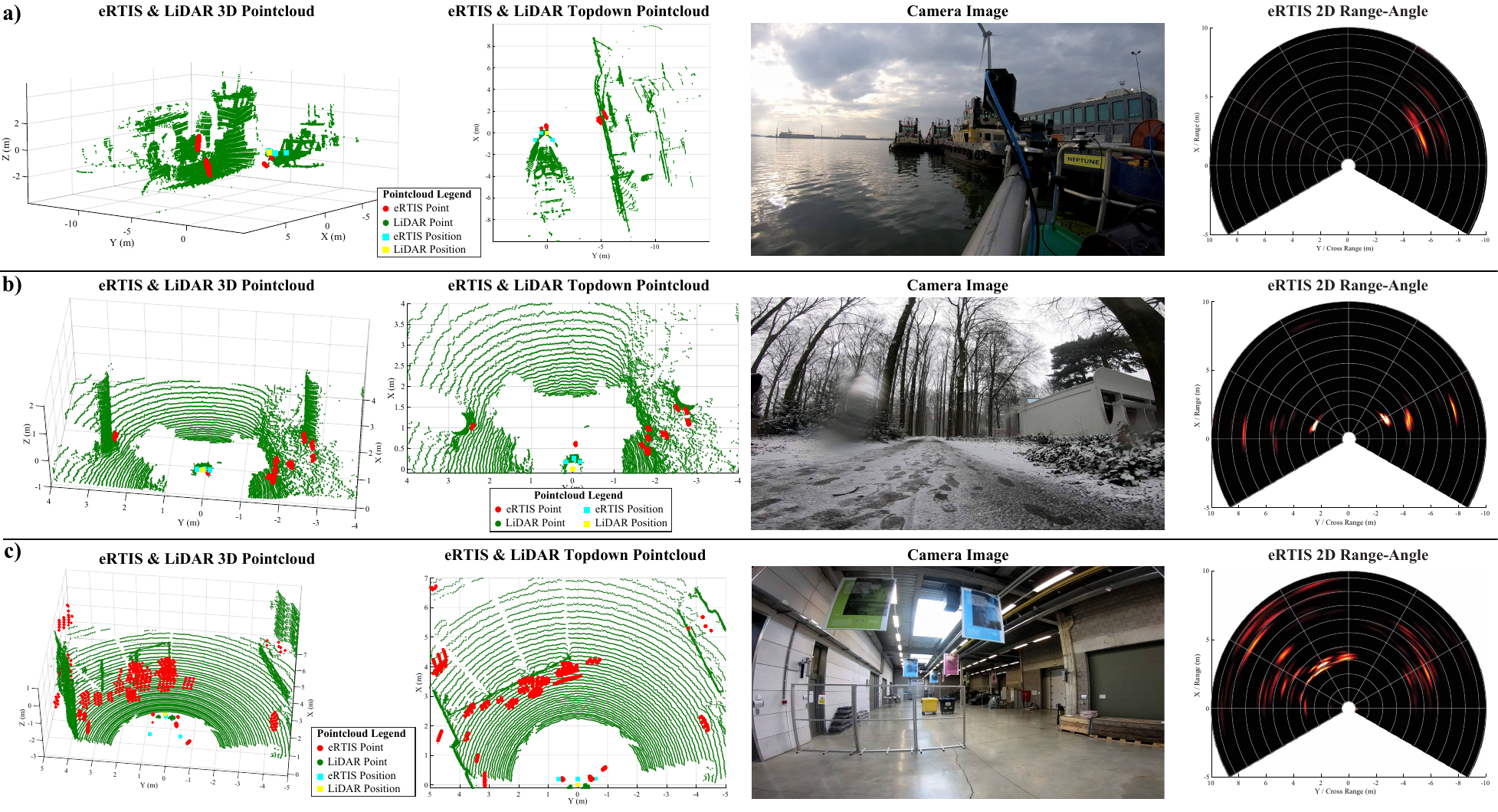}
    \caption{A single frame from the real-world experiments, showing data from three eRTIS sensors, one LiDAR, and a camera. For the eRTIS sensors both pointcloud data and a 2D horizontal range-angle was generated. (a) Mounted on a harbor mooring vessel in the Port of Antwerp. (b) Mounted on an outdoor robot. (c) Mounted on forklift in an industrial setting with glass safety barriers.}
\label{fig:datasets}
\end{figure*}

In the first domain, the sensors were mounted on a small harbor mooring vessel in the port of Antwerp in Belgium. Several maneuvers such as docking, entering and exiting a lock, passing other ships were done to further test the performance of the eRTIS sensor in this maritime domain. An example frame during docking can be seen in Fig.~\ref{fig:datasets}a. The full video of this experiment can be seen on the Cosys-lab YouTube channel~\cite{resultsvideo1}.

The second domain focused on a small mobile platform driving through outdoor snowy conditions. An example frame is shown in Fig.~\ref{fig:datasets}b. This setup was an initial test for the autonomous robot navigation project using a multi-sensor eRTIS configuration for navigation~\cite{jansen2022real}. The full video of this experiment can be seen on the Cosys-lab YouTube channel~\cite{resultsvideo2}.

The last domain focused on an industrial setting where the sensors were mounted on an autonomous forklift.

\newpage \noindent An example frame of this experiment is shown in Fig.~\ref{fig:datasets}c, where one can also see how the LiDAR sensor fails to detect the glass safety barriers while the eRTIS sensor using the acoustic medium continues to detect them. The full video of this experiment can be seen on the Cosys-lab YouTube channel~\cite{resultsvideo3}.
 
\section{Discussion and Conclusion}
\label{sec:conclusion}
\color{black}

In this work, we presented eRTIS, a modular and embedded in-air broadband ultrasound sensing system designed for operation in harsh and cluttered environments. The system combines a broadband capacitive ultrasound transducer, a 32-element configurable microphone array, and embeds a high-speed microcontroller for real-time data conditioning and acquisition with a CUDA-capable NVIDIA Jetson module for GPU-accelerated signal processing. The latter also allows for interfacing with other sensing modalities either built-in or connected through dedicated interfaces.

A key feature of eRTIS is its configurable software architecture, which exposes a high-level API to control all aspects of the measurement pipeline, including excitation signal parameters, beamforming settings, and acquisition modes. The processing chain is managed using parallel CUDA workers, transforming raw acoustic data into 2D or 3D images with low latency. The results are accessible via a TCP/UDP interface and can be directly pushed to the Robot Operating System (ROS) framework, enabling straightforward integration into robotic platforms and sensor fusion pipelines.

The system is enclosed in a sealed, anodized aluminum housing with IP-rated connectors and supports multi-device synchronization via differential signaling or in-band acoustic markers. A dedicated synchronization unit enables both simultaneous and sequenced triggering for up to six eRTIS devices.

We validated the system across three representative application domains: harbor mooring monitoring, off-road robotics, and autonomous navigation in cluttered environments. The results demonstrate that eRTIS provides robust, real-time sensing in environments where optical systems often fail, confirming the viability of embedded in-air ultrasound as an alternative sensing modality for industrial and autonomous applications.

\bibliographystyle{IEEEtran}
\bibliography{main}

\end{document}